# Freeform super-oscillatory optics for CMOS-integrated THz super-resolution imaging

Jin Chen[1#], Liang Gao[2#*], Hao Guo[1#*], Zhi Chao Chen[1,3#], Kang Jie Lin[2], Kam Man Shum[1], Ka Fai Chan[1], Chi Hou Chan[1,4*]

[1] State Key Laboratory of THz and Millimeter Waves, City University of Hong Kong, Hong Kong, 999077, China

[2] School of Information Science and Engineering, State Key Laboratory of Millimeter Waves, Southeast University, Nanjing, 210096, China

[3] Institute of Plasma Physics, Chinese Academy of Sciences, Hefei, Anhui 230031, China

[4] Department of Electrical Engineering, City University of Hong Kong, Hong Kong, 999077, China

These authors contributed equally: Jin Chen, Liang Gao, Hao Guo, Zhi Chao Chen

[*]Corresponding author: lianggao@seu.edu.cn, haoguo8@cityu.edu.hk, eechic@cityu.edu.hk

## Abstract

The diffraction limit fundamentally constrains the spatial resolution of far-field imaging systems. While near-field techniques can circumvent this limit, their inherently short working distances (WD) severely restrict practical applications. Super-oscillatory lenses (SOLs) offer a far-field alternative; however, conventional SOLs are plagued by discrete operating wavelengths, low efficiencies (<5%), and formidable trade-offs among numerical aperture, chromatic aberration, and depth of focus (DOF). Here, we introduce a nonlocal, nonlinear-curvature mechanism to design a freeform SOL that achieves ultrabroadband (0.3–1 THz), achromatic super-resolution focusing with an unprecedented efficiency of 44%. Operating at a 9-mm WD, the lens maintains a consistent sub-diffraction full-width at half-maximum (FWHM) of ~0.45λ alongside an extended DOF of ~10λ. By integrating a compact 65-nm CMOS oscillator-radiator array, we establish an advanced imaging platform capable of resolving complex 2D and 3D sub-millimeter features (down to 0.15 mm). Readily scalable to the optical regime via two-photon lithography, this freeform SOL paradigm paves the way for next-generation, high-performance integrated photonics.

## Introduction

Enhancing imaging resolution is paramount for unraveling complex structural details and discovering novel phenomena across disciplines, from life sciences to astronomy. Fundamentally, however, the conservation of electromagnetic momentum dictates that high-spatial-frequency components—which carry crucial subwavelength information—manifest as evanescent waves. Because these waves decay exponentially in the near field, they inherently elude far-field detection. Consequently, conventional optical systems are strictly bound by the classical Abbe diffraction limit (0.5λ/NA, where NA stands for numerical aperture), thereby establishing a formidable physical barrier to advancing high-resolution imaging.

For decades, extensive research has focused on circumventing the diffraction limit to achieve super-resolution imaging. These strategies broadly fall into near-field and far-field regimes. Near-field techniques—such as negative-index lenses[1-3], hyperlenses[4-6], near-field scanning optical microscopy[7], and microsphere arrays[8-11]—rely on recovering high-spatial-frequency evanescent waves to resolve fine details. However, the exponential decay of evanescent waves inherently confines the sample to be in extreme proximity to the lens, severely impeding their practical applications. Conversely, established far-field techniques, notably super-resolution fluorescence microscopy[12], require invasive fluorescent labeling and suffer from low temporal resolution. Even promising label-free approaches, such as structured illumination[13-14], necessitate multi-frame acquisition, restricting real-time observation. Recently, the rapid advancement of artificial intelligence (AI) has provided a computational alternative for super-resolution reconstruction[15]; nevertheless, the physical fidelity of such data-driven upscaling is fundamentally compromised by the "black-box" nature of neural networks[16-17]. In summary, these existing far-field and computational methods rely on intricate optical setups, external contrast agents, or algorithmic post-processing. A fundamental breakthrough in the underlying physical mechanisms—one that enables direct, single-shot far-field super-resolution—remains unsolved.

Over the past decade, the mathematical concept of superoscillation[18]—where a band-limited function locally oscillates faster than its highest Fourier component—has been successfully translated into the design of SOLs[19-39]. By achieving sub-diffraction focusing with a far-field FWHM of ~0.42λ, SOLs offer a compelling pathway for non-invasive, label-free super-resolution imaging. However, conventional SOLs that rely on diffractive elements are severely constrained by physical trade-offs. Specifically, generating highly confined electromagnetic fields (large NA) inevitably compromises the DOF. Furthermore, they suffer from exceptionally low focusing efficiencies due to the radiation of massive sidelobes, severe chromatic aberrations/narrower working bandwidth stemming from wavelength-dependent phase profiles, and restrictively short WDs. While extensive efforts have been devoted to mitigating these bottlenecks, state-of-the-art performance remains inadequate. Even in the most optimized designs[36]—where DOF and WD can exceed 10λ, the focusing efficiency plateaus at merely ~0.01%, and "achromatic" operation is strictly limited to a few discrete wavelengths. Consequently, the comprehensive performance of existing SOLs still falls significantly short of the demands for practical applications, underscoring the critical need for a paradigm shift toward high-performance, practically viable super-oscillatory optical systems.

Our recent demonstration of aberration-free THz metalenses with a wide field of view[40] has underscored the remarkable potential of gradient-index (GRIN) architectures for super-resolution imaging. Consequently, GRIN lenses represent a compelling paradigm to bypass the fundamental bottlenecks inherent in conventional diffractive SOLs. While precisely engineering complex 3D GRIN profiles with GRIN metamaterials remains a formidable nanofabrication challenge at optical frequencies, the THz regime provides an accessible and robust platform for such metamaterial design and manufacturing. Furthermore, the unique penetrability of THz waves through dielectric materials makes them exceptionally well-suited for volumetric probing.

Motivated by these dual advantages, we strategically implement our GRIN-based design methodology in the THz regime. This approach establishes an optimal platform for realizing highly efficient, ultrabroadband achromatic SOLs with a large WD and DOF, ultimately enabling robust 3D super-resolution imaging.

In this work, we propose a novel nonlinear, nonlocal GRIN lens capable of achieving high efficiency (44%), ultrabroadband achromatic super-oscillatory focusing with an extended DOF. The underlying physical mechanism relies on the complex interplay between two adjacent, identical nonlinear GRIN profiles, which synthesize the high-efficiency super-oscillatory field. To circumvent the severe fabrication challenges associated with nonlinear metamaterials, the nonlocal GRIN profile is mathematically mapped to an equivalent freeform surface, inspired by our previous microlens architectures [41]. This equivalent freeform lens is then seamlessly realized via high-precision 3D printing. Both theoretical modeling and experimental characterizations confirm that this 3D-printed sample exhibits robust, high-efficiency achromatic focusing across a massive bandwidth from 0.3 to 1 THz, maintaining a consistently narrow FWHM of ~0.45 λ and an extended DOF of ~10 λ. To translate these optical breakthroughs into a practical system, we constructed a THz 3D imaging platform by integrating a compact 65-nm CMOS-based oscillator-radiator array [42] (Fig. 1). Experimental results reveal that sub-millimeter features with a strict 0.15-mm spacing—present in both Siemens resolution charts and intricate QR codes—can be resolved with remarkable fidelity. Ultimately, this work establishes a robust paradigm for constructing cost-effective, high-performance THz imaging platforms by synergistically bridging advanced optical freeform design and scalable 3D printing.

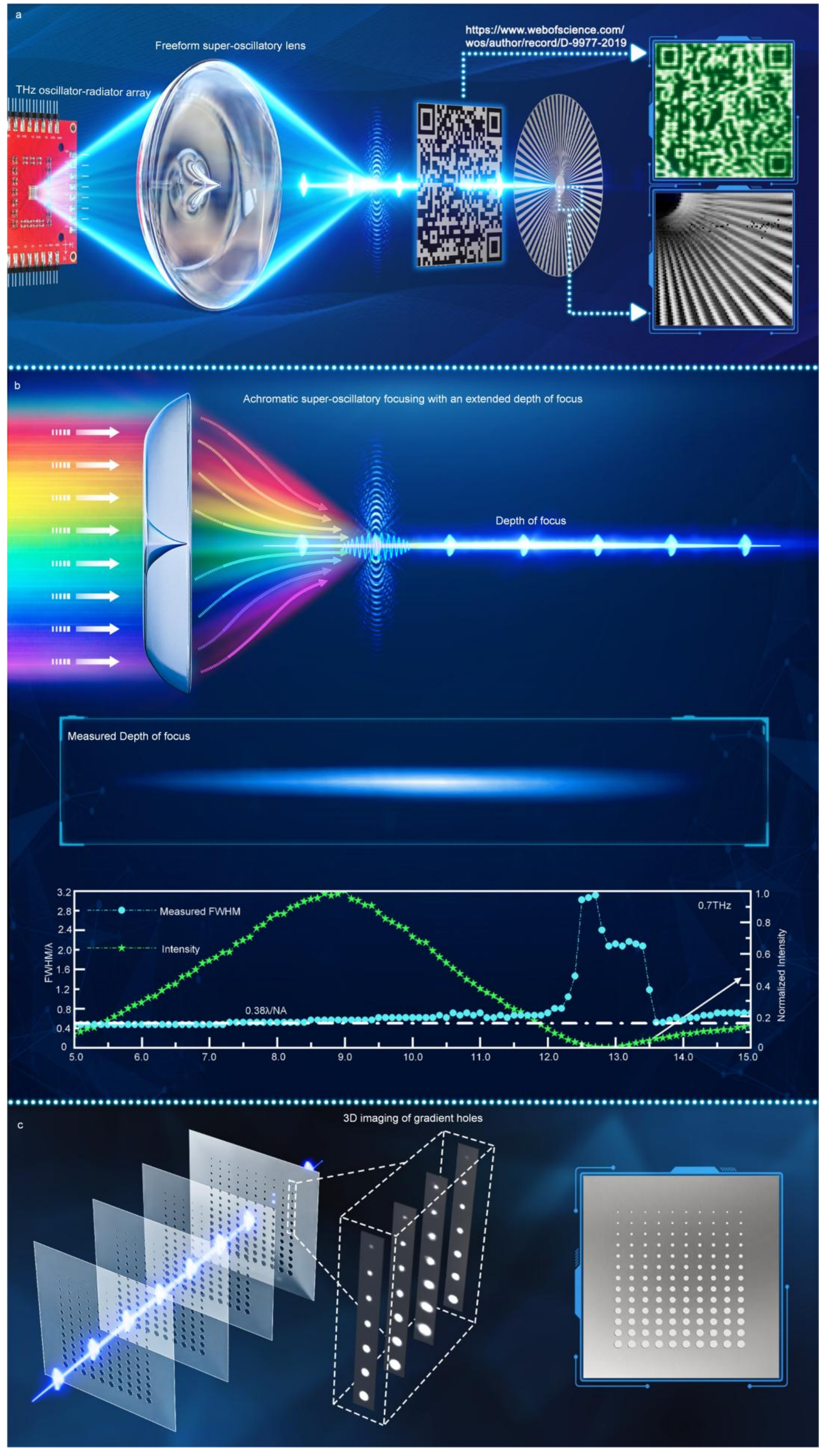
a
Freeform super-oscillatory lens
THz oscillator-radiator array
https://www.webofscience.com/
wos/author/record/D-9977-2019
b
Achromatic super-oscillatory focusing with an extended depth of focus
Depth of focus
Measured Depth of focus
Measured FWHM
Intensity
0.7THz
0.38λ/NA
FWHM/λ
Normalized Intensity
c
3D imaging of gradient holes

**Fig. 1. A super-resolution THz imaging system composed of a freeform super-oscillatory lens and THz oscillator-radiator array. a.** Integrated THz super-resolution imaging system**. b.** Achromatic super-oscillatory focusing with an extended depth of focus. **c**. 3D imaging results of gradient holes.

## Results

### Theoretical design of THz nonlocal nonlinear curvature super-oscillatory lens

In conventional linear-phase profiles, the focusing resolution is strictly dictated by NA. Consequently, achieving super-oscillatory focusing necessitates an ultra-large NA, which intrinsically severely restricts WD and limits the operational bandwidth due to severe chromatic dispersion. Alternatively, nonlinear phase profiles—most notably the cubic phase distribution—have been widely employed to generate accelerating Airy beams in both the optical and THz regimes. However, the focal spot sizes of these Airy beams remain far above the sub-diffraction threshold required for super-oscillation. Therefore, realizing super-oscillatory focusing with a novel class of nonlinear phase profiles is highly compelling.

To strike an optimal balance between spatial resolution and focusing efficiency, we investigate a fifth-power phase profile. Since the exact analytical solution for this high-order nonlinear phase is exceptionally challenging to derive directly, we propose an alternative methodology based on a nonlinear fifth-power GRIN distribution. The details of this fifth-power phase profile are given in Supplementary Note 1. Nonetheless, our initial calculations of the electric field distributions (Fig. S1) under the illumination of plane waves reveal that this basic fifth-power GRIN profile yields a FWHM of only ~0.78 λ, which is insufficient for super-oscillation. To further transcend this resolution limit, we introduce a nonlocal coupling mechanism between two adjacent nonlinear GRIN segments. By harnessing this intricate interplay, the final nonlocal nonlinear GRIN profile is mathematically formulated as follows:

$$n(r) = 1.6 - 0.6\left|\left(2\frac{r}{R} - 1\right)^5\right| \quad (1)$$

Where R is the lens aperture radius, and r denotes the radial distance from the center. For experimental validation in the THz regime, the lens is designed with R = 10 mm and a total thickness of 3 mm. To evaluate the resultant focusing performance, the corresponding electric field distributions were rigorously calculated. As demonstrated in Fig. S3, the proposed nonlocal nonlinear GRIN lens achieves robust, ultrabroadband achromatic focusing across the 0.3–1 THz spectrum. Furthermore, the extracted focusing resolution and efficiency (Fig. S4) quantitatively confirm the achievement of high-efficiency (~44%) super-oscillatory focusing with an FWHM around 0.45λ. Despite the theoretical elegance of the nonlocal nonlinear GRIN architecture, practically realizing such complex, large-aperture gradient metamaterials remains prohibitively challenging with state-of-the-art 3D printing technologies. To circumvent this formidable fabrication barrier, we draw inspiration from our previous work on microlenses[41], which established a rigorous mathematical equivalence between volumetric GRIN profiles and surface-relief freeform optics. Consequently, we strategically map the 3D nonlocal nonlinear GRIN distribution onto an equivalent 2D freeform lens topology, as schematically illustrated in Fig. 2a. To rigorously validate this geometric transformation, the spatial electric field distributions of the equivalent freeform lens on the x-z planes and x-y planes were numerically calculated and depicted in Fig. 2b and c. The results compellingly demonstrate that the freeform counterpart perfectly inherits the ultrabroadband achromatic focusing capabilities of the original GRIN design. The quantitatively extracted FWHM and focusing efficiency (Fig. 2e) confirm the robust preservation of super-oscillatory focusing, delivering an FWHM of ~0.45λ alongside an exceptional efficiency of ~50%. The extended DOF (>10λ) characteristic of this freeform architecture is comprehensively mapped in Fig. 2f. Furthermore, the evolution of the FWHM and the corresponding intensity profile of the THz needle along the propagation axis are quantitatively detailed in Fig. 2d. Crucially, the FWHM is consistently maintained below 0.38λ/NA throughout the entire DOF. This robust longitudinal confinement highlights the immense potential of the proposed nonlocal nonlinear freeform lens for high-fidelity 3D volumetric imaging. Ultimately, Fig. 2g presents a systematic benchmark comparing our design against state-of-the-art super-oscillatory lenses. The proposed freeform architecture exhibits unprecedented overall performance; most notably, it achieves a striking superiority in focusing efficiency—a critical parameter that directly dictates signal-to-noise ratio and ultimate imaging quality.

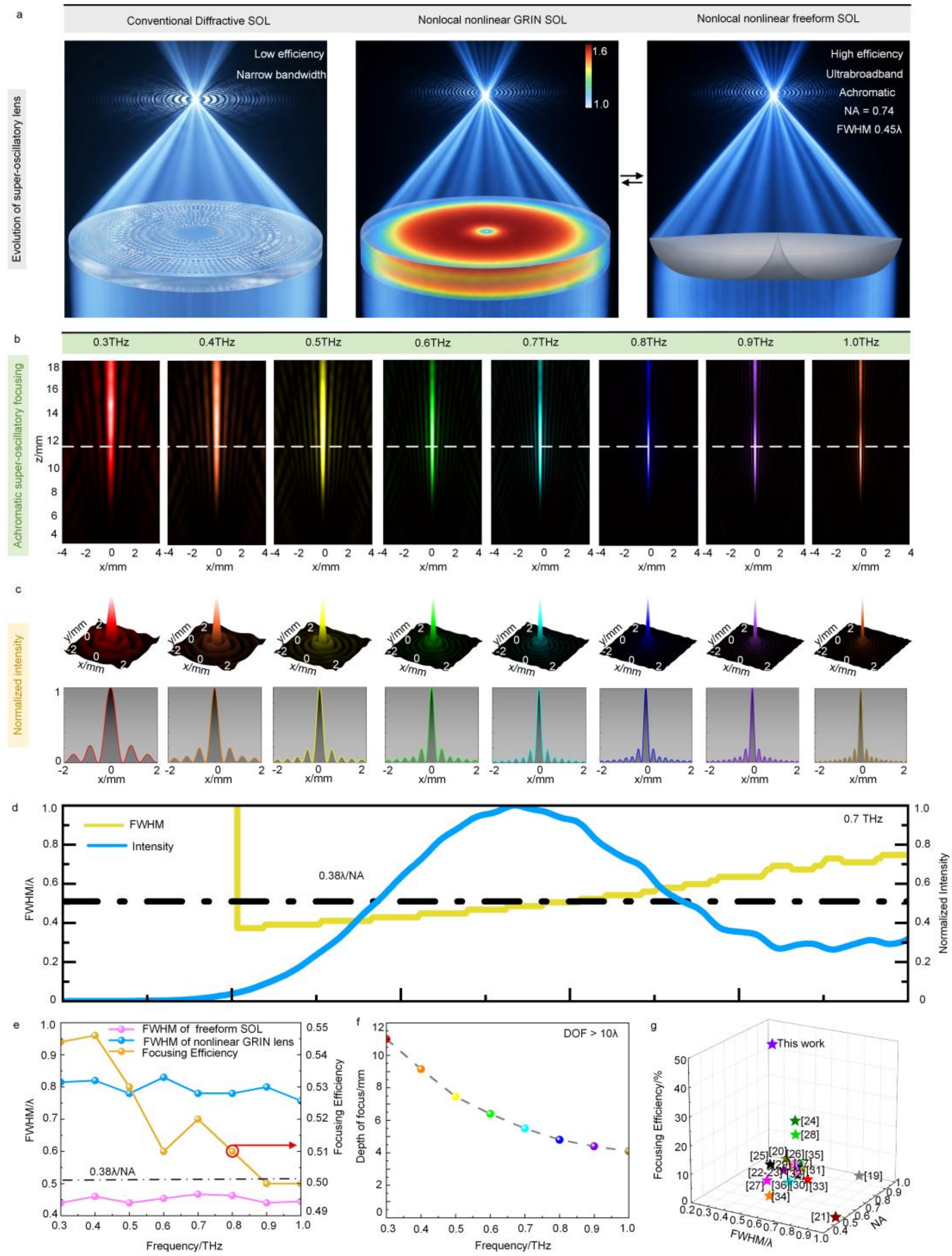

**Fig. 2. Theoretical design of a nonlocal nonlinear freeform THz SOL. a.** Evolution of the SOL from a diffractive SOL to a nonlocal nonlinear freeform SOL. **b-c**. Calculated electric-field distributions of ultra-broadband achromatic super-oscillatory focusing from 0.3 to 1.0 THz, and the extracted normalized intensity on the focusing planes. **d**. FWHM and intensity of the THz needle's main lobe along the propagation direction at 0.7 THz. **e**. Derived focusing efficiency and FWHM from Fig. 2b. **f**. Depth of focus over the 0.3–1 THz band. **g**. Overall comparison between the nonlocal nonlinear freeform lens and previous works from the perspectives of FWHM, focusing efficiency, and NA.

## Experimental demonstration of the THz nonlocal nonlinear curvature lens

A custom experimental setup was deployed to precisely map the spatial electric-field distribution and characterize the achromatic super-oscillatory behavior, as shown in Fig. S5. The measured intensity profiles in the *x-z* and *x-y* planes (Fig. 3a) unequivocally demonstrate the achievement of ultrabroadband achromatic focusing over the range 0.3 to 1.0 THz. Fig. 3b further delineates the normalized field distributions at the focal planes. Notably, the experimentally extracted FWHM values (Fig. 3c) remain remarkably stable at ~0.45λ, matching the theoretical predictions with high fidelity; slight deviations are well within typical experimental uncertainty limits. The inherent frequency dispersion of the resin used for the 3D-printed lens (Fig. S5) intrinsically compresses the effective DOF compared to ideal lossless simulations (Fig. 3d), resulting in a measured DOF that decreases from 7 mm to 3.3 mm across the operational band. Notwithstanding the dispersion effects, the preserved deep-subwavelength resolution and the millimeter-scale extended DOF rigorously confirm the exceptional suitability of this 3D-printed freeform architecture for high-resolution 3D imaging applications. The variation of the measured FWHM and the corresponding on-axis intensity profile of the THz needle along the propagation direction are presented in Fig. 3e. Notably, the FWHM remains consistently maintained at approximately 0.38λ/NA over the entire DOF, indicating the considerable potential of the proposed nonlocal nonlinear freeform lens for high-fidelity 3D imaging.

On the other hand, the measured focusing efficiency presents a monotonic negative correlation with incident frequency. As plotted in Fig. 3c, the efficiency gracefully degrades from 70% at 0.3 THz to 37% at 1.0 THz. The pronounced discrepancy between the measured and theoretical focusing efficiencies primarily originates from the extreme spatial confinement of the focal spot, which closely approaches the spatial resolution limit of the receiving probe. Because the standard probe utilizes a rectangular aperture, it exhibits a highly asymmetric spatial response near this resolution limit. As evidenced by the measured focal field profiles shown in Fig. 3a, this geometry-induced artifact artificially inflates the vertical sidelobes compared to the horizontal ones. Consequently, the focusing efficiency integrated from this distorted spatial profile inevitably underestimates the intrinsic value. Furthermore, as the frequency increases, the focal spot shrinks further toward the probe's detection limit, exacerbating this spatial convolution error and resulting in a progressively lower measured efficiency.

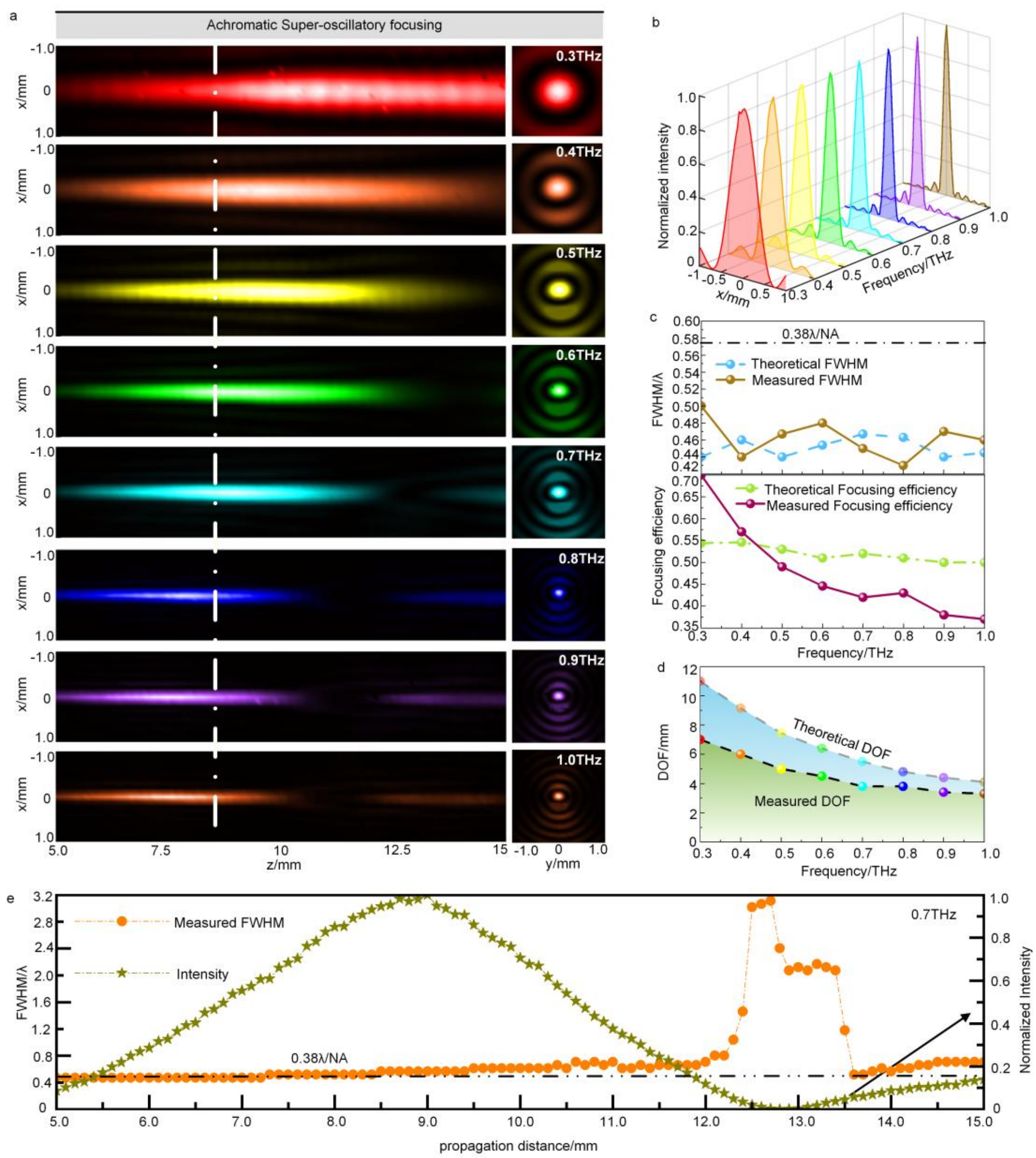

**Fig. 3 Experimental results of achromatic super-oscillatory focusing. a.** Measured results of electric field distributions in the x-z plane and x-y plane from 0.3 to 1.0 THz. **b**. Extracted electric field distributions at the focusing lines. **c**. Comparison between the measured FWHM/focusing efficiency and the theoretical one. **d**. Comparison between the measured DOF and the theoretical one. **e**. Measured FWHM and intensity of the THz needle's main lobe along the propagation direction at 0.7 THz.

## A compact THz imaging system of a nonlocal nonlinear curvature lens and an oscillator–radiator array

Integrating the proposed nonlocal nonlinear curvature lens with the oscillator-radiator array, we developed the compact THz imaging system depicted in Fig. S7. In this configuration, the array emits *y*-polarized THz radiation at 0.658THz, which is subsequently shaped by the nonlocal nonlinear curvature lens into a super-oscillatory beam with an extended depth of field (DOF) to illuminate the sample. Subsequently, the transmitted THz signals are captured by a receiving probe. For 2D image acquisition, the sample is mounted on a motorized stage for continuous *x*-*y* raster scanning, and the recorded intensity data are mathematically post-processed to reconstruct the final images.

Since the intrinsic sidelobes of conventional super-oscillatory beams often introduce artifacts that degrade the resolution of complex patterns, we employed a USAF 1951 resolution test chart, a Siemens star resolution chart, and a QR code (with minimum feature sizes of 0.2 mm) to rigorously demonstrate the superior imaging fidelity of our system (Fig. 4). Specifically, the USAF 1951 resolution chart and Siemens star resolution chart were fabricated by selectively etching a thin aluminum sheet, while the QR code was lithographically patterned into the inner layer of a thin PCB panel (see the optical photograph in Fig. S6).

Despite the influence of pronounced background artifacts in the raw imaging data, the fine structural details of both the USAF 1951 resolution target, the Siemens star resolution chart, and the embedded QR codes are distinctly resolved. Remarkably, the raw THz image of the 0.2-mm-feature QR code is immediately machine-readable; even without any algorithmic post-processing, it can be successfully decoded by a standard smartphone to the URL of Web of Science. This direct readability unambiguously demonstrates the robust imaging capability of the proposed engineered lens when resolving intricate patterns.

To further suppress background artifacts and elevate image quality, a customized post-processing pipeline—comprising deconvolution, stripe filtering, and axial constraints—was applied to the raw dataset. The corresponding details are provided in the Methods. As depicted in Fig. 4a-c, this procedure substantially mitigates the artifacts, yielding a striking improvement in image clarity. The processed data exhibits markedly superior contrast compared to the original raw image. Therefore, the minimum resolvable feature sizes for the USAF 1951 chart and the QR codes are firmly determined to be 0.15 mm and 0.2 mm, respectively, and we establish that the reliable spatial resolution of the proposed compact THz imaging system for complex structural patterns is 0.2 mm. Furthermore, the imaging results obtained using the Siemens star resolution target demonstrate that the designed THz imaging system can resolve not only samples with a constant feature size, such as QR codes, but also samples featuring progressively varying feature dimensions, which is rather difficult for a super-oscillatory lens.

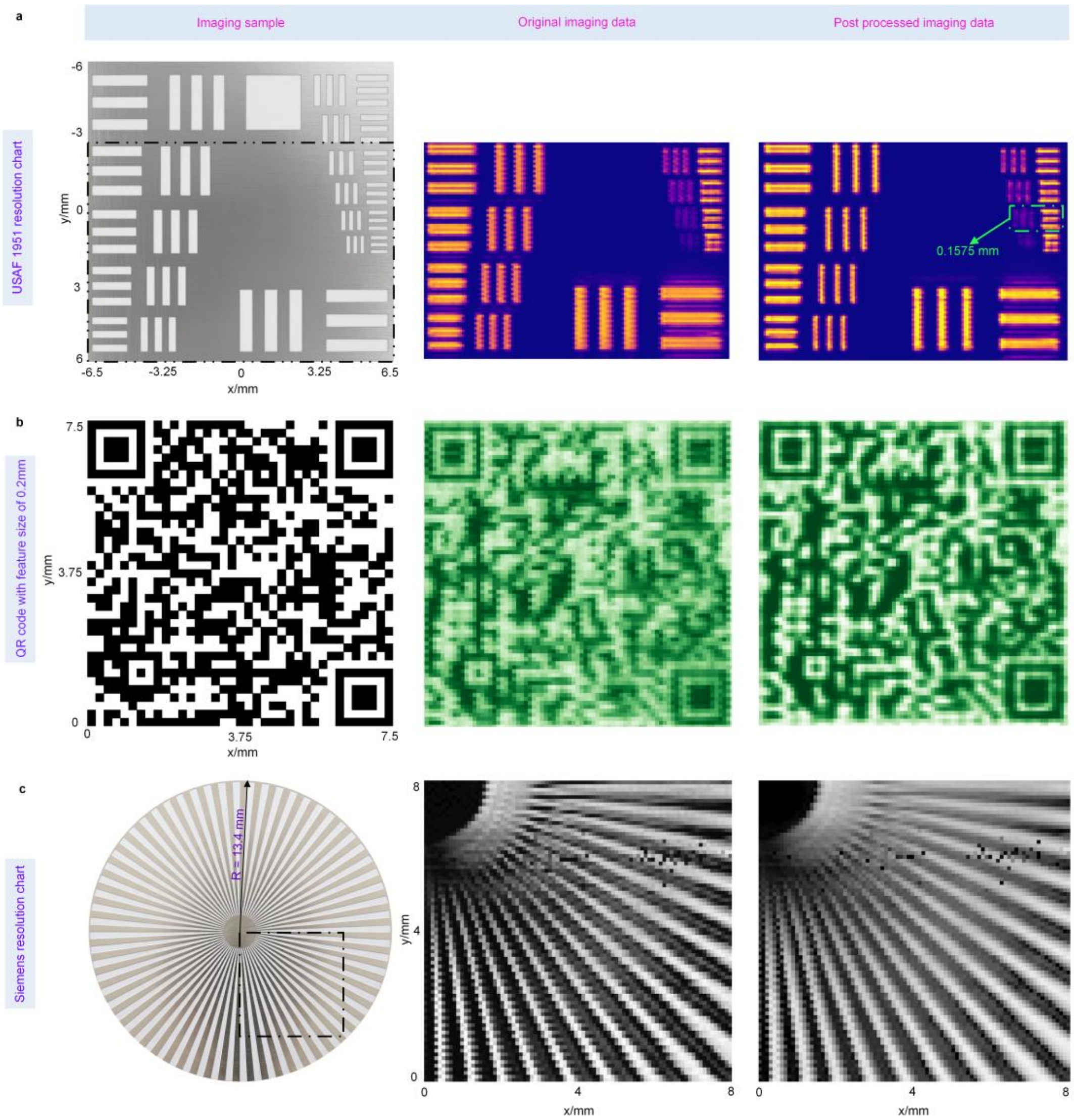

**Fig. 4. Imaging results of the CMOS integrated THz imaging system**. **a-c**. Imaging samples and imaging results of the USAF 1951 resolution chart, QR codes with a feature size of 0.2 mm, and a Siemens resolution chart.

Finally, leveraging the extended depth of focus (DOF) of 4 mm (at 0.658 THz) provided by the nonlocal nonlinear curvature super-oscillatory lens, we explored the system's volumetric 3D imaging capabilities. An aluminum sheet containing an array of gradient-radius holes was employed as the test sample and scanned at various axial positions within the DOF. As depicted in Fig. 5, the overall tomographic imaging quality remains highly stable across all four depth planes. While the absolute diffraction limit—resolving the smallest hole with a 0.15-mm radius—is exclusively achieved at the exact focal center, indicating that the highest resolution of the USAF 1951 resolution chart and small holes remains at the same level. The macroscopic imaging resolution is robustly maintained throughout the entire 4 mm DOF, confirming the extended volumetric resolving power of our system.

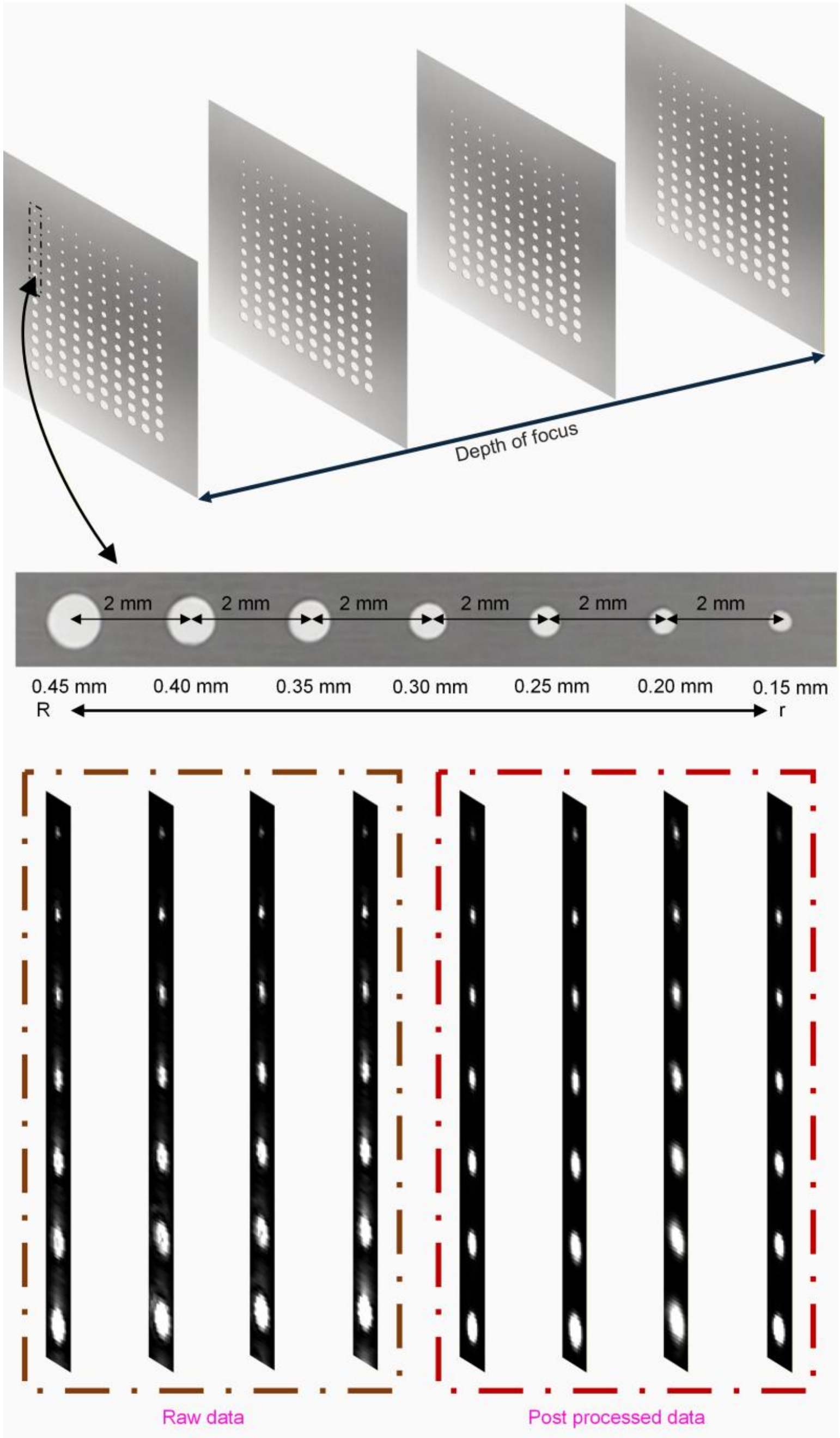

**Fig. 5.** Imaging results of gradient holes at different depths of DOF

## Discussion

In summary, we have demonstrated an integrated THz super-resolution imaging platform by combining a freeform super-oscillatory lens with a 65-nm CMOS-based oscillator–radiator array. Through nonlinear-curvature design and nonlocal wavefront engineering, the lens achieves efficient, ultrabroadband achromatic super-resolution focusing over 0.3–1 THz, with a focusing efficiency of 50%, an extended DOF of approximately 10 wavelengths, and a nearly invariant FWHM of ~0.45λ at a 9-mm WD. Upon integration with the oscillator–radiator array, the system achieves a spatial resolution of 0.15 mm on a USAF 1951 resolution target and reliably resolves complex QR-code patterns with invariant feature dimensions, together with Siemens resolution targets exhibiting radially graded feature sizes. These results demonstrate the robustness of the proposed platform in resolving both discrete and continuously varying subwavelength features. Ultimately, 3D imaging of gradient holes is also accomplished. By integrating freeform super-oscillatory optics with standard CMOS technology, this work provides a compact, efficient, and scalable paradigm for high-performance THz imaging systems, with potential extension to optical frequencies via two-photon lithography.

## Methods

The observed image was modeled as the convolution of the target transmittance with the nonlocal nonlinear curvature super-oscillatory lens' Point Spread Function (PSF), corrupted by additive noise:

$$m(x,y) = (p * t)(x,y) + n(x,y) \tag{2}$$

where $m$ is the measured image, $t$ is the target transmittance, $p$ is the PSF, and $n$ is noise. Periodic stripe artifacts were first suppressed in the Fourier domain using narrow-band notch filtering:

$$m_f(x,y) = Q\{m\}(x,y) = F^{-1}\{K(u,v)H(u,v)\} \tag{3}$$

Where $K = F\{m\}$ and $H(u,v)$ is the notch-filter transfer function, $Q$ is the frequency-domain stripe filtering. To match the sampling grid of the PSF, the destriped image was upsampled by

$$N = round\left(\frac{l_{img}}{l_{psf}}\right) \tag{4}$$

defines the upsampling ratio, $l_{img}$ and $l_{psf}$ denote the scanning steps of the image and the PSF, respectively, with $l_{img} < l_{psf}$ in our setup. And a bicubic interpolation operator $U_N$ is applied to the destriped image at factor $N$ so that the equivalent pixel size after upsampling matches that of the PSF, yielding

$$m_{\uparrow}(x,y) = U_N\{m_f\} \tag{5}$$

$m_{\uparrow}$ is the upsampled image. The corresponding high-resolution degradation model is

$$m_{\uparrow}(x,y) = (p * t_{\uparrow}) + n_{\uparrow} \tag{6}$$

where $t_{\uparrow}$ is the high-resolution transmittance to be recovered and $n_{\uparrow}$ is the equivalent noise on the upsampled grid. The high-resolution transmittance was then estimated using TV-regularized Richardson–Lucy deconvolution. The update rule was[43-46]

$$\hat{t}^{(k+1)} = \frac{\hat{t}^{k}}{1-\lambda\nabla\cdot\frac{\nabla\hat{t}^{(k)}}{\left\|\nabla\hat{t}^{k}\right\|+\eta}}\bullet\left[p*\frac{m_{\uparrow}}{p*\hat{t}^{(k)}+\varepsilon}\right] \quad (7)$$

Where $\hat{t}^{(k)}$ is the estimate $t_{\uparrow}$ at the $k$-th iteration, $\tilde{p}*$ is the flipped PSF, λ is the TV regularization weight, and $\xi$ and $\eta$ are numerical stabilization constants. To suppress deconvolution-induced oblique sub-pixel artifacts, an axial projection correction was applied:

$$\hat{t} \leftarrow \hat{t} - \beta\left(E_{+}^{\mathrm{T}}E_{+}\hat{t} + E_{-}^{\mathrm{T}}E_{-}\hat{t}\right) \quad (8)$$

Where $E_{+}$ and $E_{-}$ are diagonal finite-difference operators and $\beta$ is the step size. Finally, the result was downsampled to the native image grid by $N\times N$ area averaging:

$$\hat{t}(x,y) = \frac{1}{N^2}\sum_{(i,j)\in\Omega_{N(x,y)}}\hat{t}_{\uparrow}(i,j) \quad (0)$$

where $\Omega_{x,y}$ denotes the $N\times N$ region on the high-resolution grid corresponding to the original pixel $(x,y)$. The down-sampling operator is denoted as $E_{N}$. $R_{p,\lambda,k*}$ is TV-regularized Richardson–Lucy deconvolution, and $B_{\beta}$ is the axial projection constraint. Therefore, the complete processing pipeline can be written as the following operator cascade:

$$\hat{t} = E_{N}\circ B_{\beta}\circ R_{p,\lambda,k*}\circ U_{N}\circ Q(m) \quad (10)$$

**Data availability**

The data that support the findings of this study are presented in the paper and the Supplementary Information file.

**Code availability**

The data that support the findings of this study are available from the corresponding authors upon reasonable request.

**Acknowledgments**
This research was supported in part by the Research Grants Council of the Hong Kong under the General Research Fund (grant CityU 11214123).

Author information

These authors contributed equally: Jin Chen, Liang Gao, Hao Guo, Zhi Chao Chen
Authors and affiliations.

**State Key Laboratory of THz and Millimeter Waves, City University of Hong Kong, Hong Kong, 999077, China**
Jin Chen, Hao Guo, Zhi Chao Chen, Kam Man Shum, Ka Fai Chan, & Chi Hou Chan
**School of Information Science and Engineering, State Key Laboratory of Millimeter Waves, Southeast University, Nanjing, 210096, China**

Liang Gao, Kang Jie Lin

**Institute of Plasma Physics, Chinese Academy of Sciences, Hefei, Anhui 230031, China**

Zhi Chao Chen

**Department of Electrical Engineering, City University of Hong Kong, Hong Kong, 999077, China**
Chi Hou Chan

Contributions

C.H.C., L. G, and H. G initiated the plan and supervised the entire study. C.H.C. and J.C. conceived the idea of this work. J. C. carried out the lens design and implemented the theoretical analysis. L. G designed the THz radiator array. H. G and K. J. L. completed the testing of the THz radiator array. J.C., H.G., K.S.M., and K.F.C. designed the experimental setup. J.C. and H.G. performed the imaging experiments. Z. C. C. performed the post-processing of the original imaging data. J.C. analysed the data and prepared the manuscript with input from all authors. C.H.C., H. G and L. G reviewed and edited the manuscript.

Corresponding author

Correspondence to Liang Gao, Hao Guo, and Chi Hou Chan.

**Ethics declarations**

Competing interests

The authors declare no competing interests.

Freeform super-oscillatory optics for CMOS-integrated terahertz super-resolution imaging

## Supplementary Note 1: Theoretical analysis of local and nonlocal nonlinear GRIN lens

The specific local fifth-power nonlinear GRIN profile is written as:

$$n(r) = 1.6 - 0.6\left(\frac{r}{R}\right)^5 \tag{1}$$

In which R is the radius of the lens, and r represents the distance from the center of the lens. The spatial electric field distributions under incident plane wave illumination were systematically calculated, as illustrated in Fig. S1.

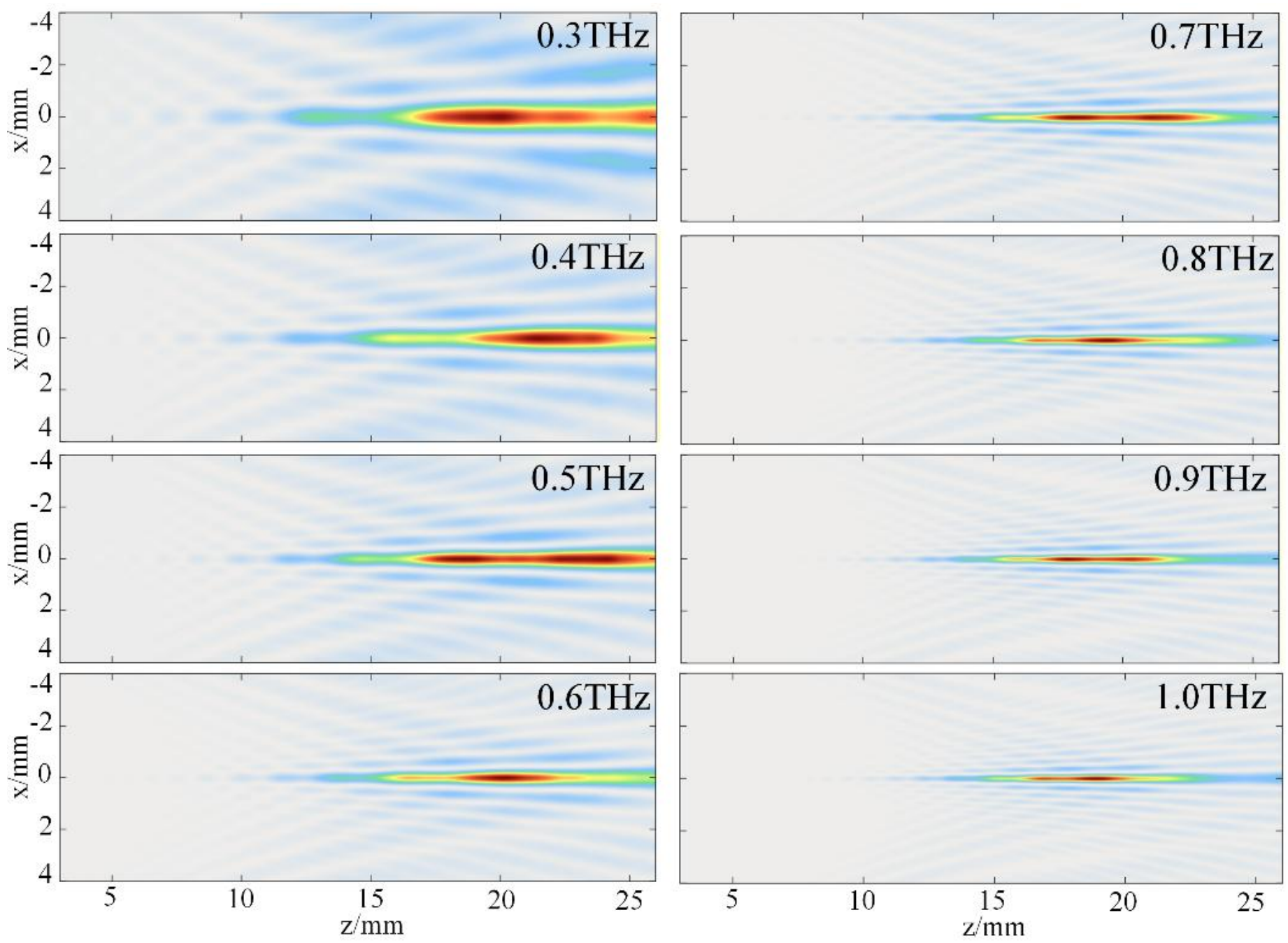

**Fig. S1** Calculated electric field distributions of the local nonlinear GRIN lens from 0.3 to 1.0 THz**.**

Furthermore, the extracted FWHM and focusing efficiency are illustrated in Fig. S2. Quantitative analysis reveals a distinct physical trade-off: while the subwavelength resolution is remarkably preserved across the spectrum (maintaining an FWHM of 0.78λ), the focusing efficiency exhibits a pronounced frequency-dependent degradation, dropping from 70% to 50%.

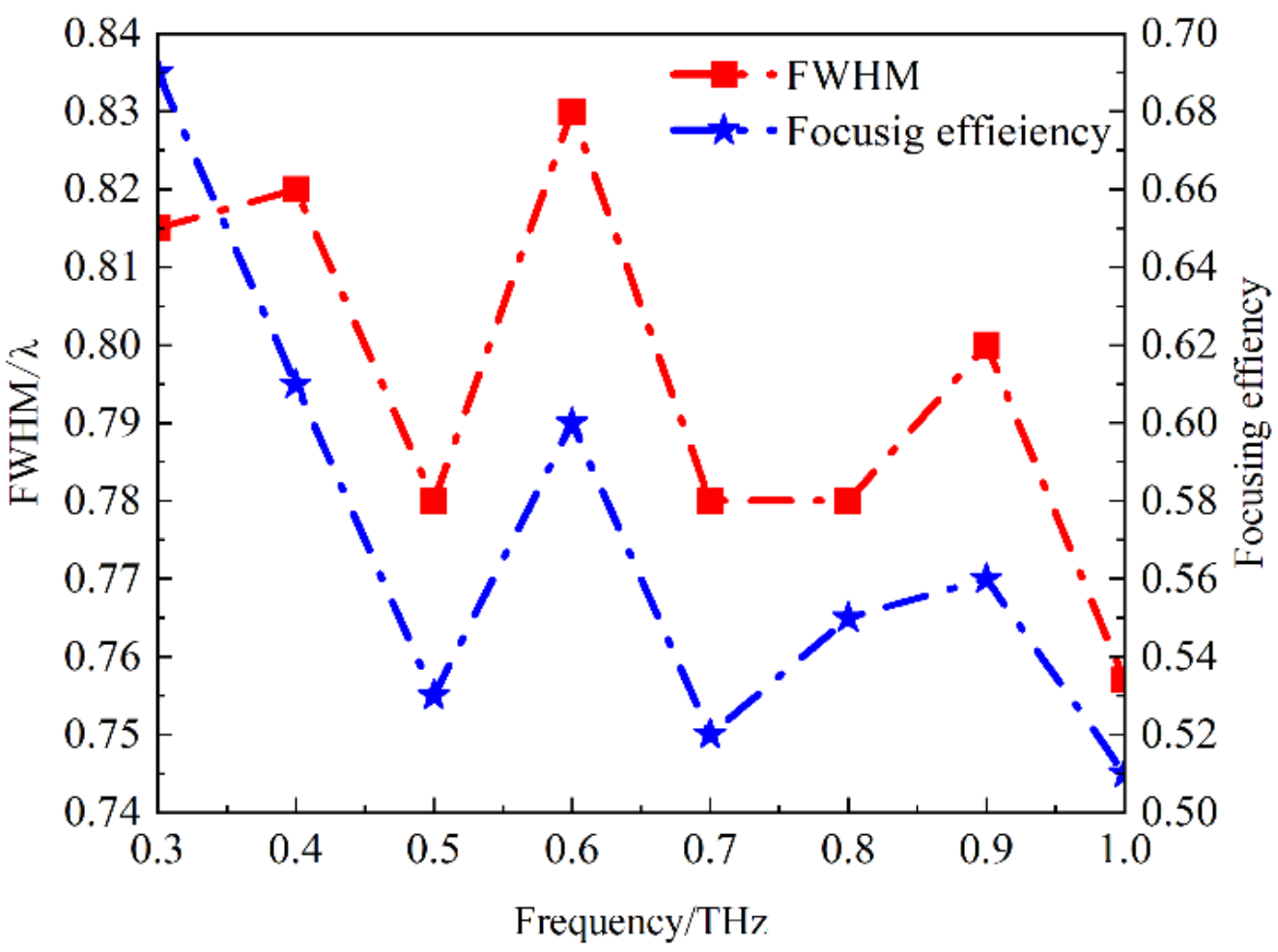

**Fig. S2** Derived FWHM and focusing efficiency.

To further enhance the focusing resolution of the nonlinear GRIN lens, a nonlocal mechanism is introduced between the adjacent nonlinear GRIN profiles, and the resulting nonlocal nonlinear GRIN profile is written in Equation 1 in the main text.

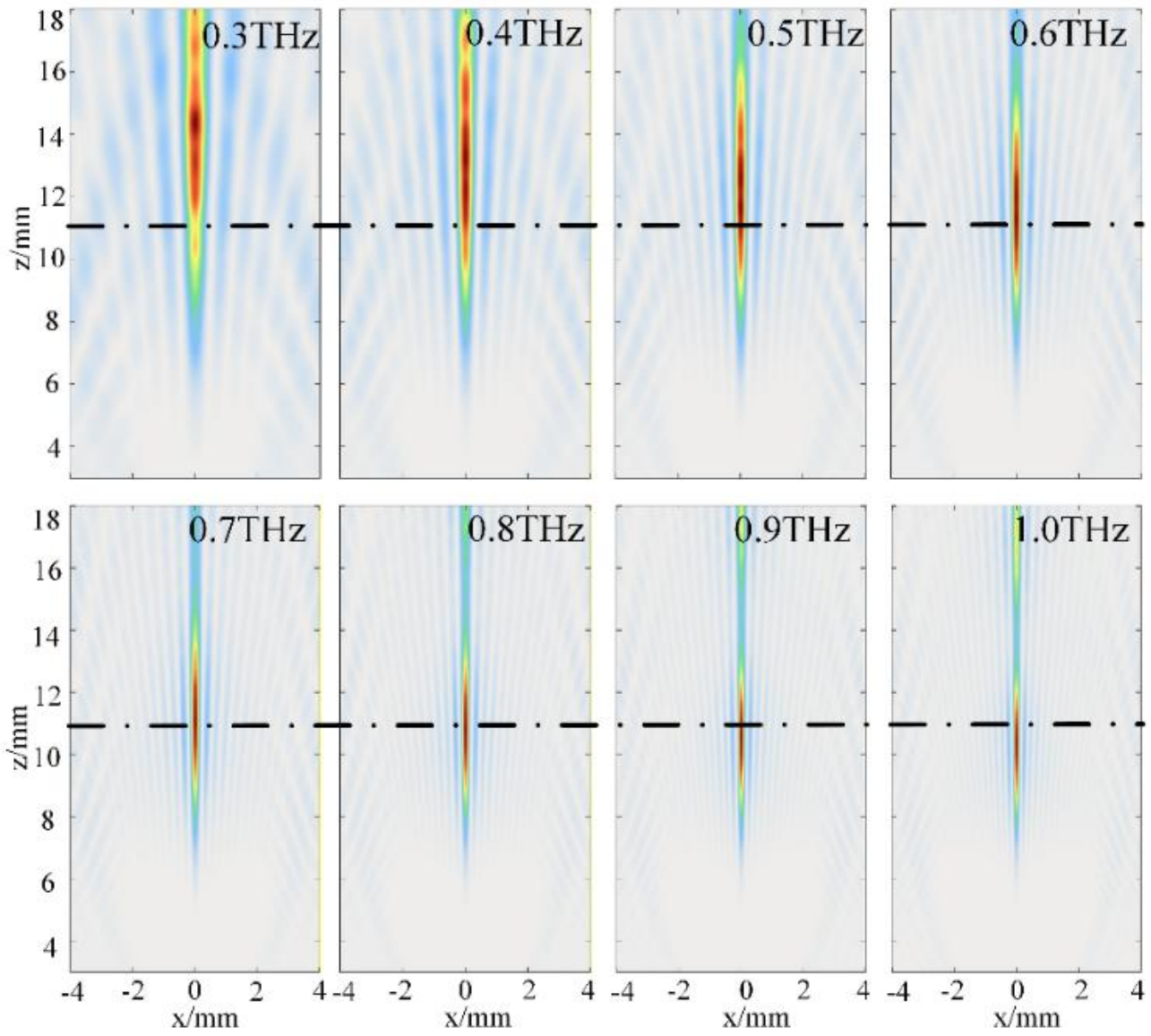

**Fig. S3** Calculated electric field distributions of the nonlocal nonlinear GRIN lens from 0.3 to 1 THz.

Figure S3 illustrates the simulated spatial electric field distributions of the nonlocal nonlinear GRIN lens across the broadband 0.3–1 THz spectrum. To rigorously evaluate its performance, the quantitatively extracted FWHM and focusing efficiency are plotted in Fig. S4 and benchmarked against the local nonlinear GRIN architecture. Strikingly, the nonlocal mechanism yields a tightly confined focal spot with an FWHM of ~0.45λ. This represents a substantially higher resolving power compared to the local design (0.78λ) and fundamentally surpasses the theoretical resolution limit typically predicted for super-oscillatory focusing (0.56λ). Crucially, despite this remarkable 1.7-fold enhancement in spatial resolution, the system successfully circumvents the traditional resolution-efficiency trade-off; the focusing efficiency remains highly comparable to that of the local GRIN lens across most of the bandwidth, except for the lowest frequency channels (0.3 and 0.4 THz).

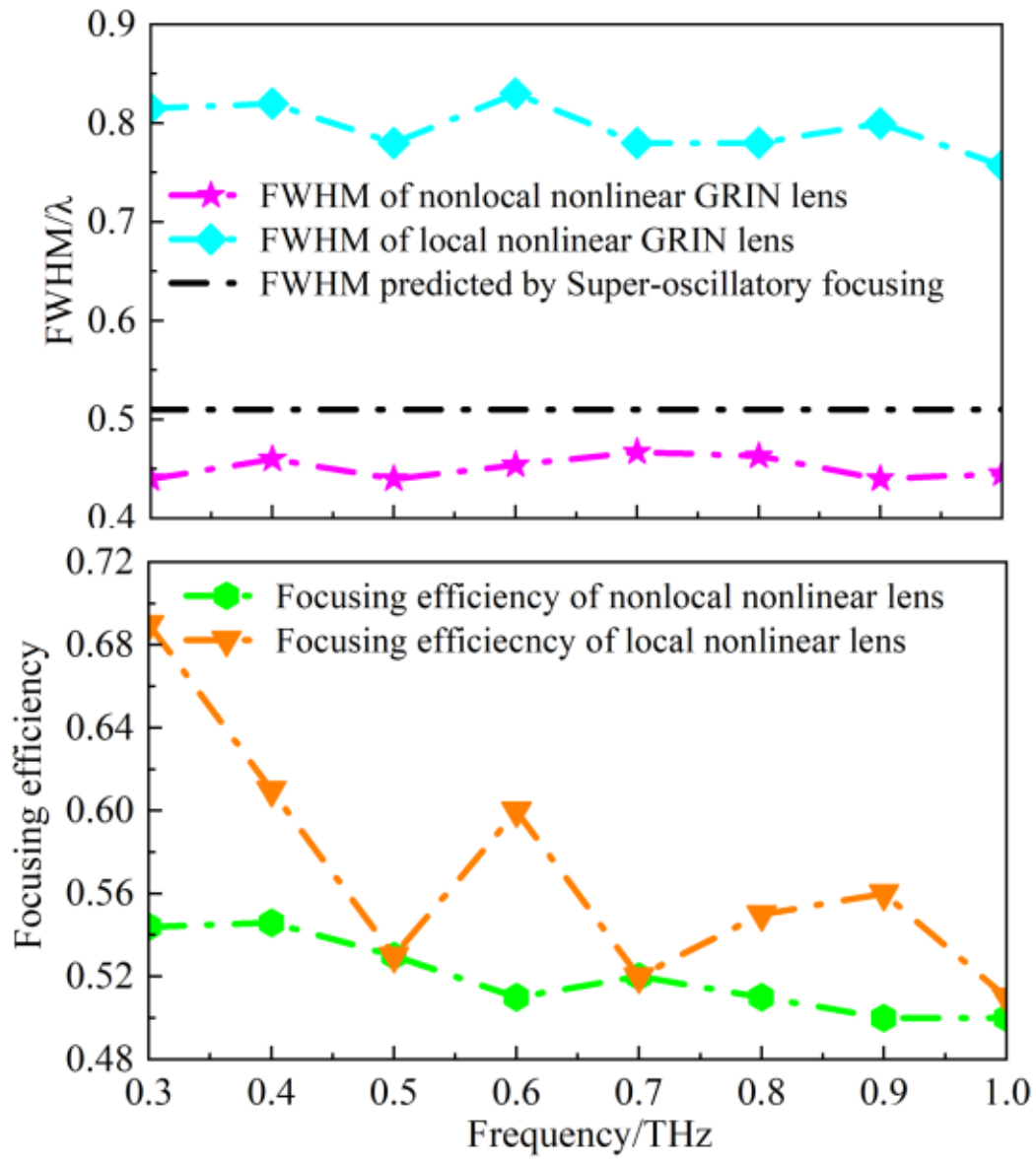

**Fig. S4**. Comparison of FWHM and focusing efficiency between the nonlocal nonlinear GRIN lens and the local linear one**.**

## Supplementary Note 2: Details of the experimental setup and measured results

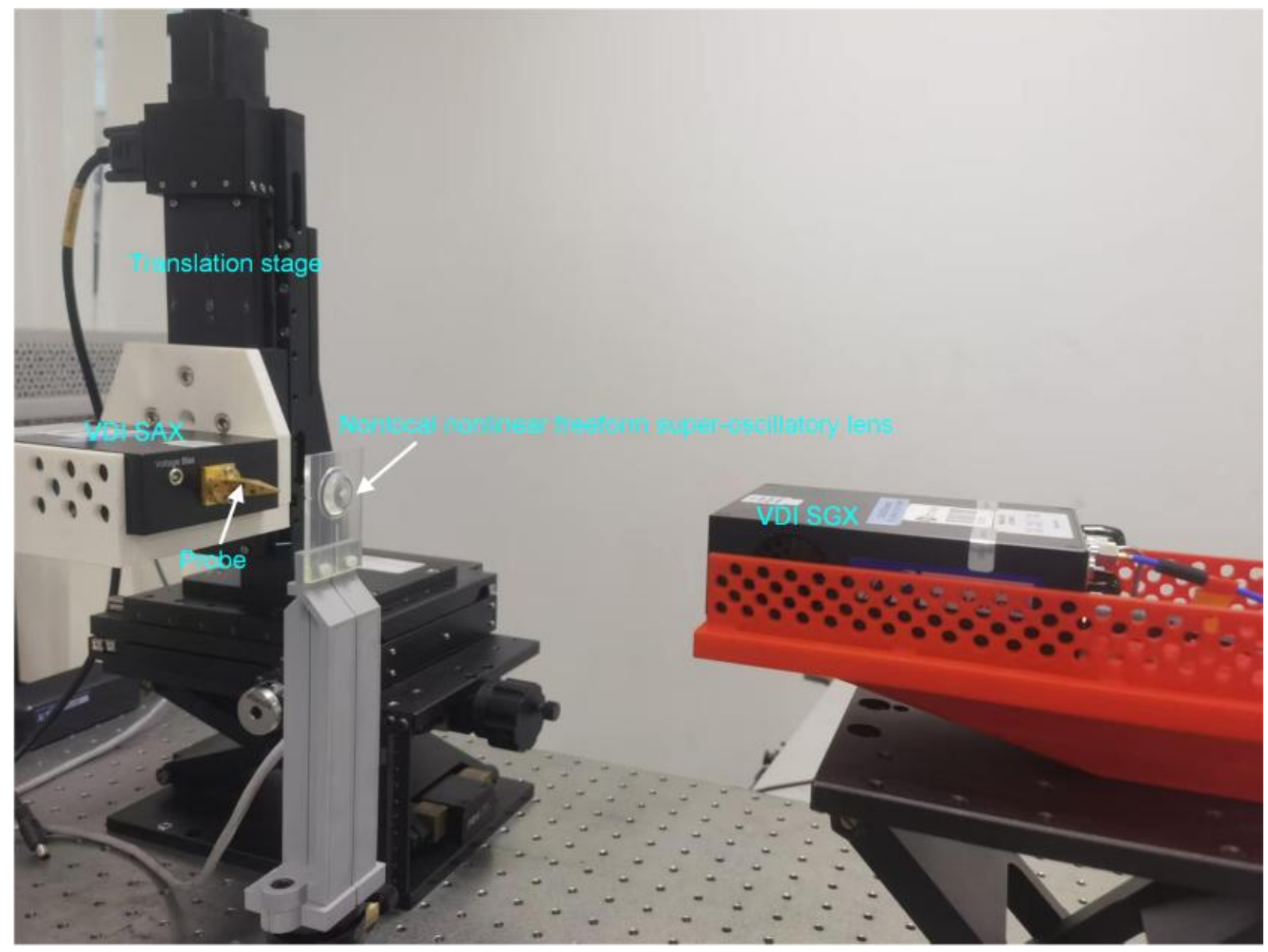

**Fig. S5**. Experimental setup for measuring super-oscillatory focusing**.**

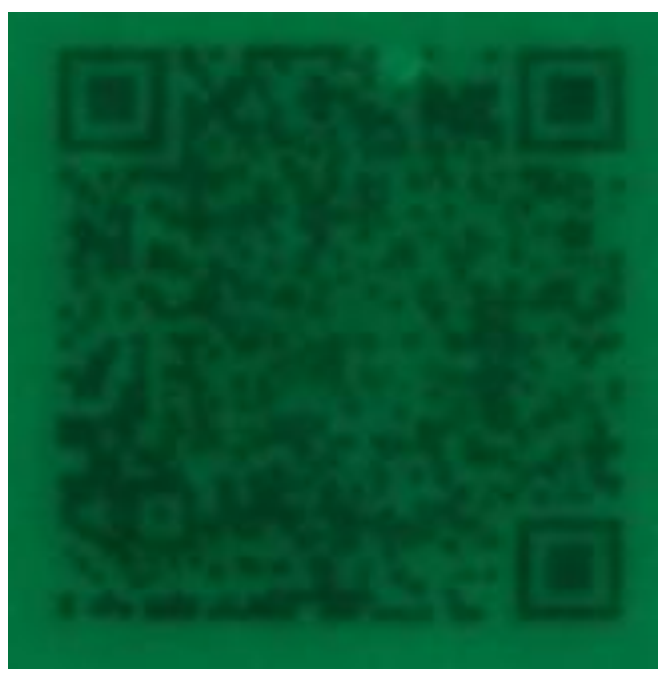

**Fig. S6**. Optical photograph of OR codes**.**

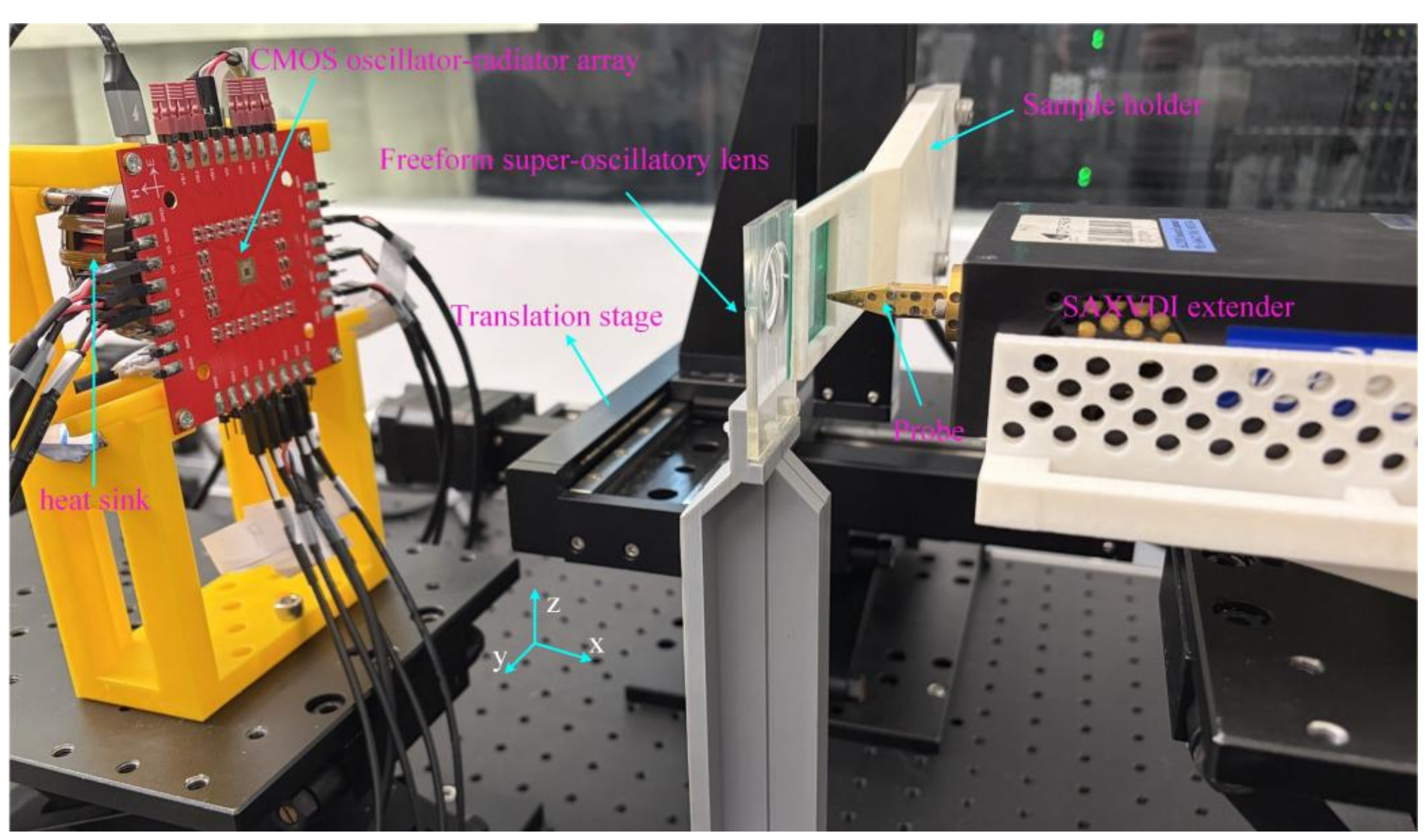

**Fig. S7**. Experimental setup for CMOS integrated THz imaging system**.**